\documentclass[final,1p,times]{elsarticle} 

\usepackage{graphicx}
\usepackage{amssymb} 
\usepackage{amsthm} 

\journal{Nuclear Physics A} 

\begin{document}

\begin{frontmatter} 

\title{ALICE results on quarkonia}

\author{Enrico Scomparin (for the ALICE\fnref{col1} Collaboration)}
\fntext[col1] {A list of members of the ALICE Collaboration and acknowledgements can be found at the end of this issue.}
\address{INFN Torino, Italy}


\begin{abstract} 
The ALICE experiment has measured quarkonia production in pp and Pb-Pb collisions at the CERN LHC, in the rapidity ranges $|y|<0.9$ and $2.5<y<4$. Quarkonia are considered to be a sensitive probe of deconfinement, and a detailed differential study of their yields can give important information on 
the properties of the medium created in heavy-ion collisions. In this paper, we will mainly discuss the centrality dependence of the J/$\psi$ nuclear modification factors, as well as their $p_{\rm T}$ and $y$ dependence in bins 
of centrality, which will be then compared to theoretical models. Preliminary results on the J/$\psi$ elliptic flow and on $\psi(2S)$ production will also be shown. 

\end{abstract} 

\end{frontmatter} 


\section{Introduction}

The suppression of quarkonium states in heavy-ion collisions was proposed long ago~\cite{Mat86} as a clean signature of a formation of a Quark-Gluon Plasma (QGP). In addition to the color screening mechanism proposed in~\cite{Mat86} as the main source of suppression, it became soon clear that other effects may contribute, as dissociation in cold nuclear matter~\cite{ZCo11} and/or in the hot confined medium~\cite{Cap08}. Also initial state effects, such as nuclear shadowing and initial state parton energy loss are expected to play a role~\cite{Vog00}. On the experimental side, lots of data are now available at SPS~\cite{Ale05,Arn07} and RHIC~\cite{Ada11,Ada12} energies, and a quarkonium (Q) suppression beyond cold nuclear matter effects (the so-called ``anomalous suppression'') was detected,  
either by studying the ratio $\sigma_{Q}/\sigma_{Drell-Yan}$ or the nuclear modification factor $R_{\rm AA}$, for different collision systems and 
as a function of the centrality and of the quarkonium kinematic variables~\cite{Bra11}. It is commonly accepted that the observed anomalous suppression is likely to be related to the formation of a deconfined state in nuclear collisions.
In this situation, the first ALICE results~\cite{Abe12} brought in a considerable surprise, showing for the J/$\psi$ clear hints of a smaller suppression, at low transverse momentum, compared to observations at lower energies. Partonic transport models~\cite{Zha11,Liu09}, as well as statistical generation models~\cite{And11}, interpret such an effect as being due to a (re)generation of J/$\psi$ along the collision history and/or at hadronization, favoured by the large $c\overline{c}$ multiplicity occurring at LHC energy.
A confirmation of this interpretation clearly requires a comprehensive study of the J/$\psi$ yield in nuclear collisions, with an emphasis on its dependence on kinematic variables. Such a study requires a sizeable Pb-Pb statistics, which was indeed collected by the ALICE experiment at the end of 2011, and results from this data taking are the main object of this paper. An accurate set of pp data is a fundamental pre-requisite for a study of nuclear effects on any observable. Moreover, important QCD-related topics can be investigated by the study of quarkonium hadro-production in elementary collisions. We will therefore also summarize the main outcomes of ALICE pp studies on quarkonia.

\section{ALICE experiment and data taking conditions}

ALICE is the dedicated heavy-ion experiment at the CERN LHC. Details on the experimental set-up can be found in ~\cite{Aam08}. The measurement of quarkonia production is carried out in the central barrel ($|y|<0.9$) through their $e^+e^-$ decay, while at forward rapidity ($2.5<y<4$)\footnote{In the ALICE reference frame, the muon spectrometer covers a negative pseudorapidity interval ($-4<\eta<-2.5$). Since in pp and Pb-Pb the physics is symmetric with respect to $y$=0, we drop the negative sign when quoting rapidity.} the $\mu^+\mu^-$ decay is studied in the muon spectrometer. 

For the measurement in the central barrel, the data acquisition is triggered by a minimum bias condition (MB), essentially defined using information from the two innermost planes of the Inner Tracker System  and from two scintillator arrays (VZERO) placed at forward rapidities (-3.7$<$ $\eta$ $<$-1.7,  2.8$<$ $\eta$ $<$5.1). In Pb-Pb, several centrality classes can be further defined at the trigger level by means of thresholds on the total signal amplitude in the VZERO. For the measurement in the forward muon spectrometer, one can require, in addition to the MB condition, one or two (depending on the data taking) candidate muons to be detected by the muon triggering system. 

The analysis presented in this paper is based, for Pb-Pb collisions at central rapidity, on a combination of 2010 data, taken with a MB trigger, and 2011 data  where, in addition, centrality selections corresponding to very central events ($\sim 0-10$\%) and semi-central events ($\sim 10-50$\%) were performed. The integrated luminosity is $L_{\rm int}\sim 15\, \mu{\rm b}^{-1}$. 
The Pb-Pb forward rapidity results refer to 2011 data ($L_{\rm int}\sim 70\, \mu{\rm b}^{-1}$), and were collected by requiring a dimuon trigger, which can select both opposite- and like-sign pairs. 
In order to properly define $R_{\rm AA}$, one needs a sample of pp data collected at the same energy ($\sqrt{s}=2.76$ TeV) and in the same kinematic domain of Pb-Pb data. Such data were taken in 2011, collecting 
$L_{\rm int}\sim 1.1\, {\rm nb}^{-1}$ with a MB trigger for the central rapidity analysis and $L_{\rm int}\sim 20\,  {\rm nb}^{-1}$ with a single-muon trigger for the forward rapidity analysis.
Finally, pp data at $\sqrt{s}=7$ TeV, the top energy of the 2010/2011 LHC data taking, were also collected and analysed, the results presented here referring to integrated luminosities up to $\sim$100 nb$^{-1}$. 

\section{Data analysis}

\begin{figure}[htbp]
\begin{center}
\includegraphics[width=0.37\textwidth]{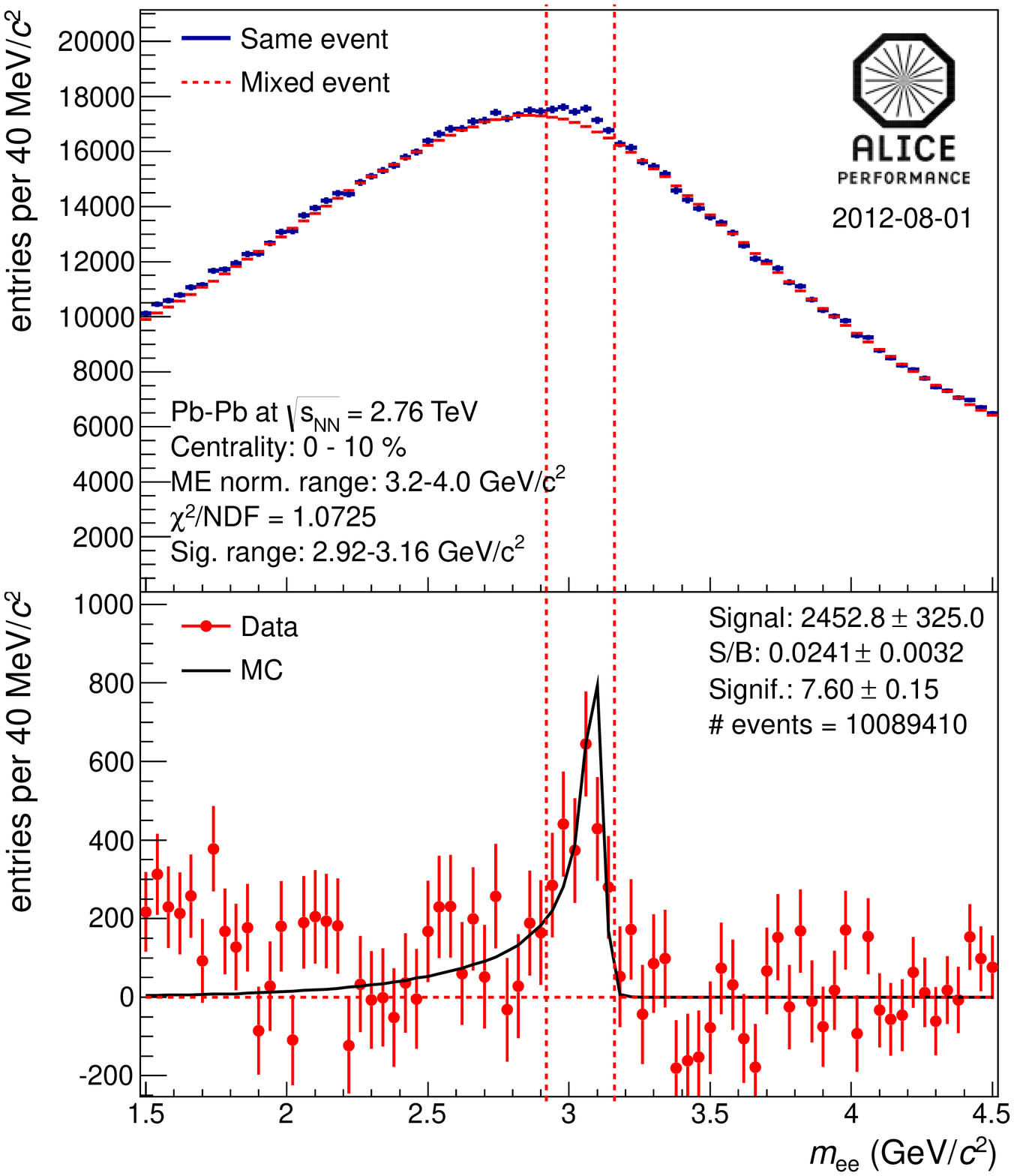}
\includegraphics[width=0.48\textwidth]{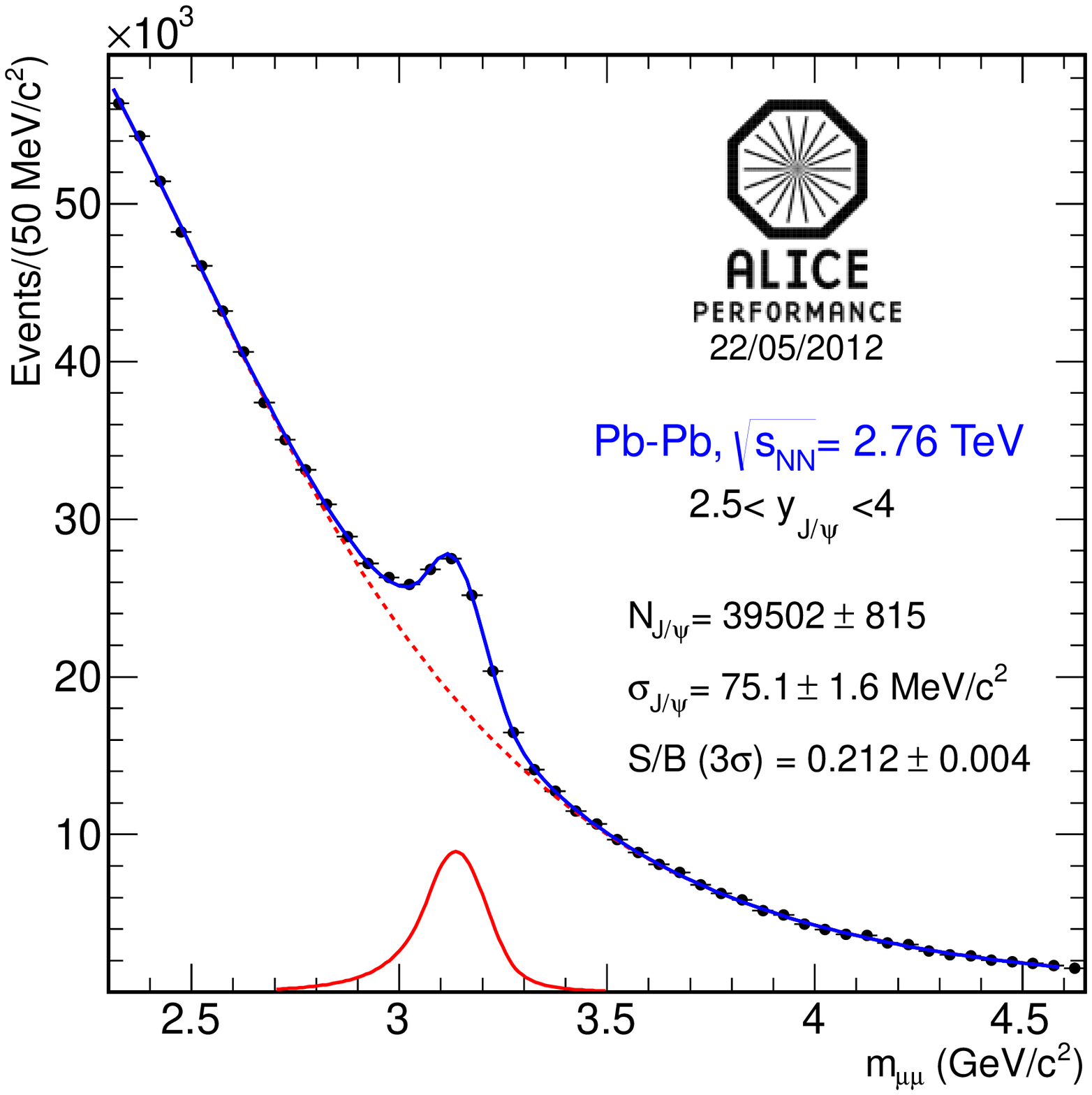}
\end{center}
\vskip -0.5truecm
\caption{Left: invariant mass spectrum for Pb-Pb (0-10\% central) opposite-sign electron pairs and background calculated using a mixed event technique (top), background subtracted spectrum with MC expected shape superimposed (bottom). Right: $\mu^+\mu^-$ invariant mass spectrum. The result of the fit around the J/$\psi$ invariant mass region is superimposed.}
\label{fig:1}
\end{figure}

The signal extraction is performed by means of an analysis of the invariant mass spectra, for both $e^+e^-$ and $\mu^+\mu^-$ decay modes. In the electron channel a large combinatorial background is present, and its size is evaluated using an event mixing technique. The mixed event sample is then normalized to the opposite sign (OS) spectrum in the mass region beyond the J/$\psi$ and subtracted out. The number of signal events is obtained by bin counting in the mass range 2.92-3.16 GeV/$c^2$, the factor extrapolating to the full mass range being obtained via Monte Carlo (MC) simulation. Fig.~\ref{fig:1} (left) shows the invariant mass spectra before and after background subtraction for the case where the latter is larger, i.e. central Pb-Pb collisions. 

In the muon channel, the J/$\psi$ signal is extracted by means of a fit to the OS mass spectrum, using a phenomenological shape for the background, and a Crystal Ball shape for the signal. The result, for centrality integrated Pb-Pb collisions is shown in Fig.~\ref{fig:1} (right), and the total statistics amount to $\sim 4\cdot 10^4$ signal events (the $\psi(2S)$ signal contribution has no practical effects on the J/$\psi$ yield, and was neglected in this fit).

The product $A\times\epsilon$ of acceptance times efficiency was calculated via MC using as input simulated pp or Pb-Pb events. Typical values amount to $\sim 5-10\%$ in the $|y|<0.9$ region (electrons)~\cite{Ars12} and $15-30$\% for $2.5<y<4$ (muons), the latter range of values mainly depending on the choice of the trigger threshold and on the request for 1 or 2 matched muons between tracking and trigger. For the muon channel in Pb-Pb, the calculation was done embedding MC J/$\psi$ inside real MB collisions. The centrality dependence of $A\times\epsilon$ in Pb-Pb collisions is found to be weak, not exceeding $~\sim 8$\% between central and peripheral events in the muon analysis, and even less in the electron analysis.

For pp collisions, the J/$\psi$ yields are then normalized to the integrated luminosity, calculated starting from the number of collected MB events. The absolute J/$\psi$ cross sections are then obtained relatively to the MB cross section, obtained via a van der Meer scan. The total systematic uncertainty on the cross section measurements ranges, depending on the data taking, between 8 and 12\% in the muon channel and between 14 and 18\% in the electron channel.

For Pb-Pb collisions, the results are given in terms of the nuclear modification factor $R_{\rm AA}$, calculated normalizing the J/$\psi$ yield to the corresponding pp cross section times the nuclear overlap function, derived from a Glauber calculation. On these analyses, an important source of (global) systematic uncertainty is related to the total error on the pp reference cross section. It is by far dominant for the electron analysis ($\sim$26\%) and represents a sizeable contribution also in the muon analysis where for example it accounts for about 50\% of the total uncertainty for the integrated $R_{\rm AA}$. 

\section{Results}

We first briefly summarize the main results obtained by ALICE in the study of pp collisions. These include the calculation of the inclusive production cross sections at $\sqrt{s} = 7$ and 2.76 TeV. At 7 TeV the differential cross sections ${\rm d}^2\sigma/{\rm d}y{\rm d}p_{\rm T}$ were obtained at both central and forward rapidity~\cite{Aam11}, where they showed a good agreement with results from the other LHC experiments (see~\cite{Aam11} for a comparison) in the common kinematic ranges, and were unique for what concerns the low-$p_{\rm T}$ region at midrapidity.
At $\sqrt{s} = 2.76$ TeV a differential study was possible only in the forward region, while at midrapidity a $p_{\rm T}$-integrated cross section was obtained, due to the rather small MB integrated luminosity~\cite{Abe122}.

Other interesting studies performed in pp include the first study at LHC energy of the inclusive J/$\psi$ polarization. The results, published in~\cite{Abe12p}, show that J/$\psi$ production is essentially unpolarized up to $p_{\rm T}= 8$ GeV/$c$. The study of this observable has critical implications for the theoretical understanding of quarkonium production, since models were quite unable to reproduce coherently the results obtained at lower energy. However, first comparisons of ALICE results with NLO NRQCD calculations show an encouraging agreement in the explored $p_{\rm T}$ range~\cite{But12}. Another study unique to ALICE is the determination of the dependence of the inclusive J/$\psi$ yield on the charged multiplicity measured at central rapidity. The results, published in~\cite{Abe12m}, show a linear increase of the yield with charged multiplicity, an observation still waiting for a satisfactory explanation from theory. Finally, at midrapidity, ALICE has provided a first measurement of the fraction of J/$\psi$ from b-decays at low $p_{\rm T}$, which also leads to an estimate of the total $b{\overline b}$ cross section~\cite{Abe12b}.

\begin{figure}[htbp]
\begin{center}
\includegraphics[width=0.48\textwidth]{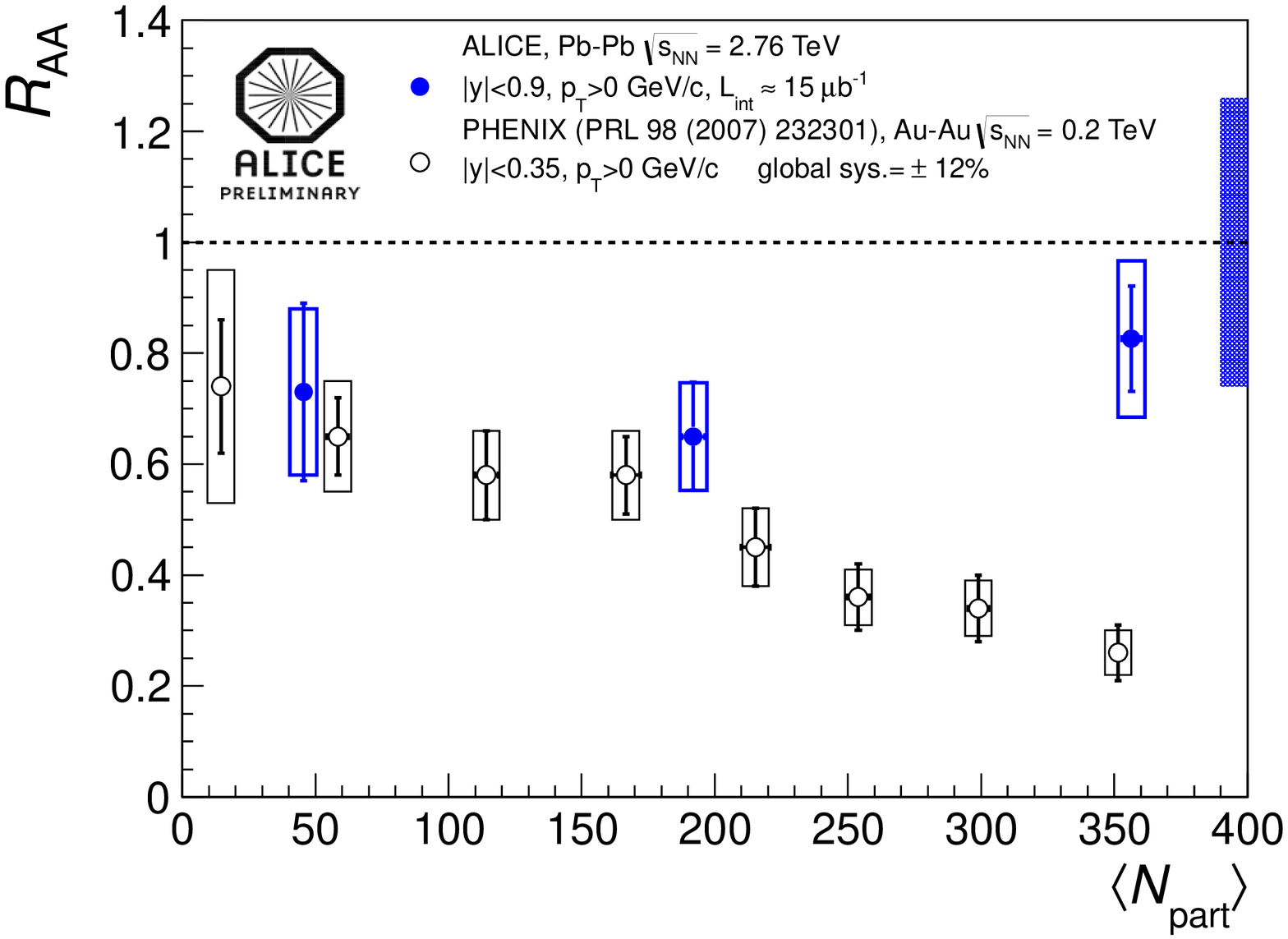}
\includegraphics[width=0.48\textwidth]{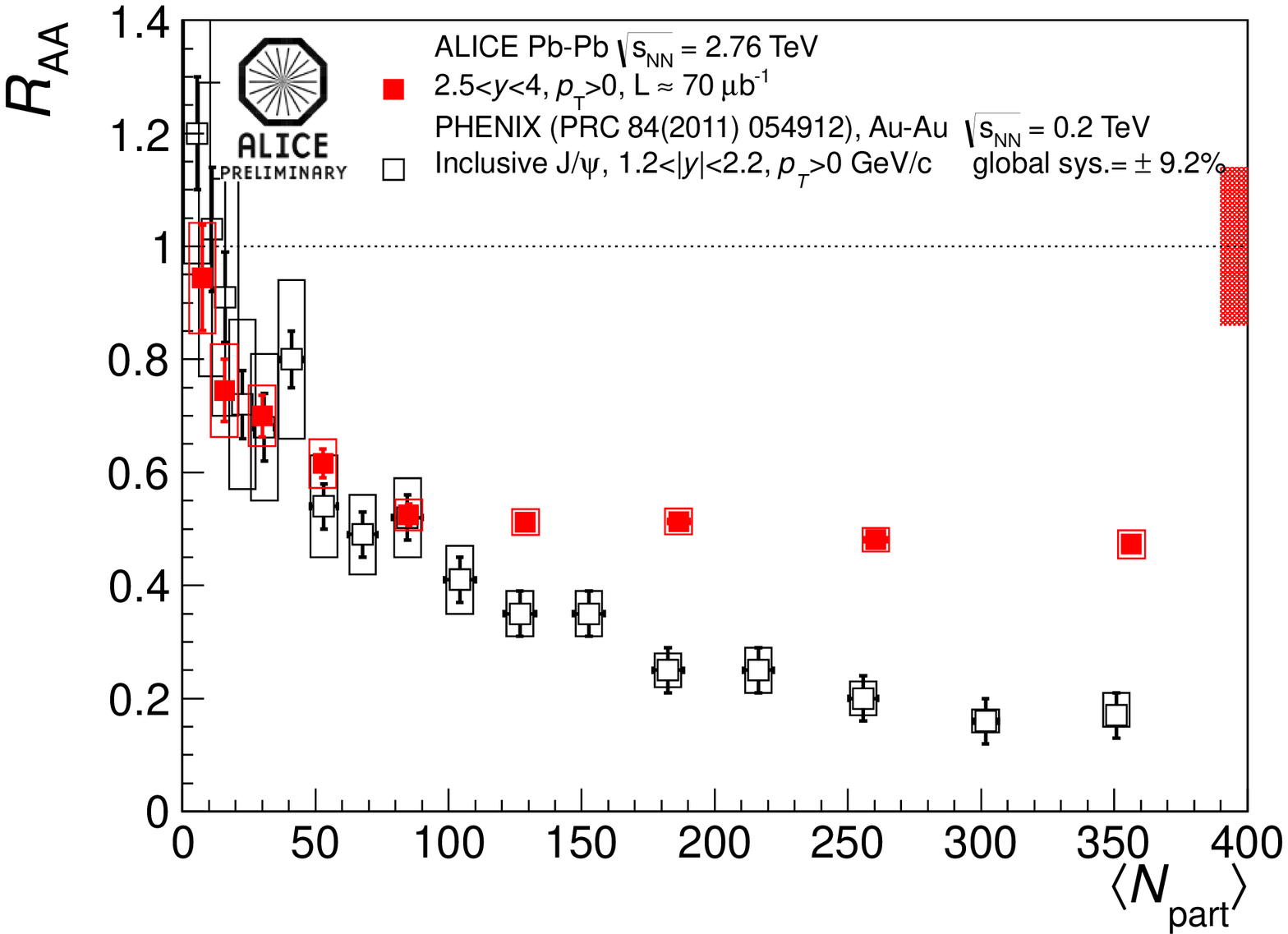}
\end{center}
\vskip -0.5truecm
\caption{Inclusive J/$\psi$ $R_{\rm AA}$ vs the number of participant nucleons $N_{\rm part}$, compared with results from PHENIX. The left plot corresponds to the central rapidity results, the right plot to forward rapidity. The boxes around the points represent the uncorrelated systematic uncertainties, while the filled boxes on the right correspond to global systematic uncertainties.}
\label{fig:2}
\end{figure}

Moving to Pb-Pb collisions, we present in Fig.~\ref{fig:2} the centrality dependence of the inclusive J/$\psi$ nuclear modification factor (see also \cite{Sui12, Arn12, Ars12}), compared with results obtained by PHENIX. While lower energy results show an increasing suppression moving towards central collisions, ALICE results clearly indicate a saturation of the suppression at both central and forward rapidity. While no final conclusion can be drawn on the relative size of the suppression measured in ALICE in the two rapidity ranges, mainly due to the large normalization error corresponding to the reference pp cross section, there is a clear evidence for a smaller suppression at LHC with respect to RHIC energy. Partonic transport models which include a (re)generation process for J/$\psi$ due to the recombination of $c\overline c$ pairs along the history of the collisions indeed predict such a behaviour~\cite{Cap08,Zha11,Liu09}, the smaller suppression at the LHC being due to the larger $c\overline c$ pair multiplicity which compensates the suppression from color screening in the deconfined phase. A similar behaviour is obtained by the statistical model~\cite{And11}, where the J/$\psi$ yield is completely determined by the chemical freeze-out conditions and by the abundance of $c\overline c$ pairs.

\begin{figure}[htbp]
\begin{center}
\includegraphics[width=0.48\textwidth]{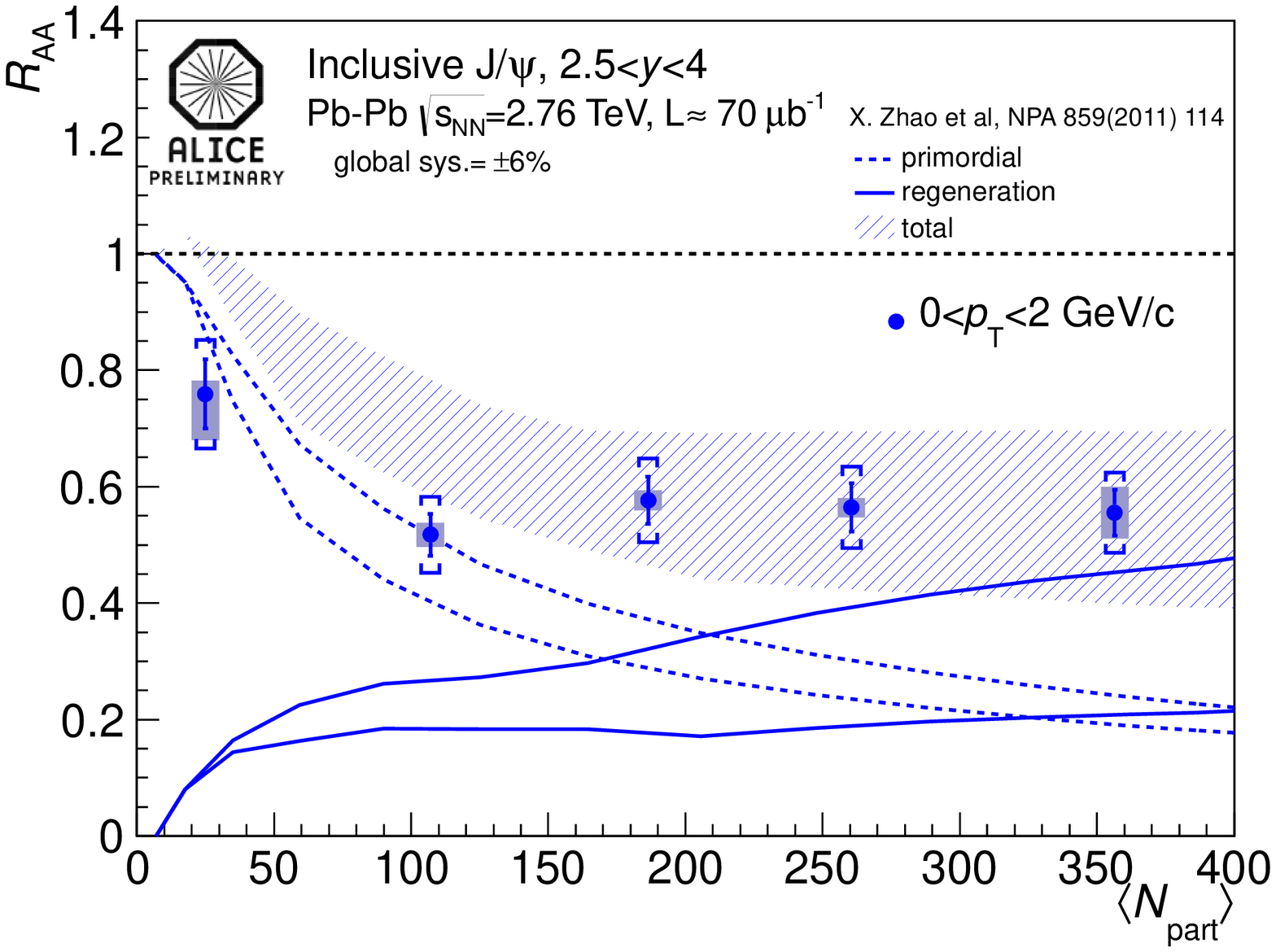}
\includegraphics[width=0.48\textwidth]{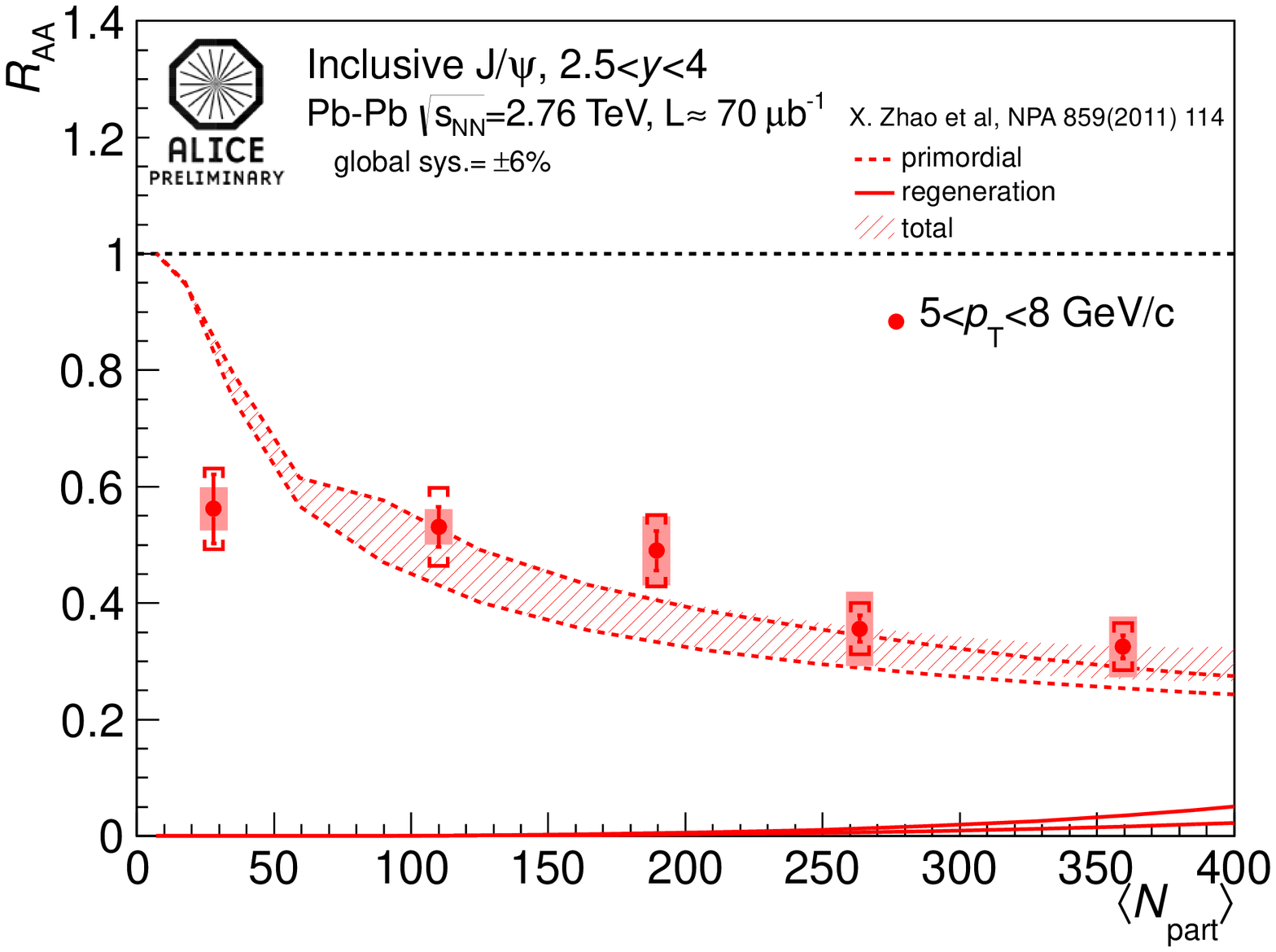}
\end{center}
\vskip -0.5truecm
\caption{$R_{\rm AA}$ vs centrality for two transverse momentum bins (0$<$ $p_{\rm T}$ $<$2 and 5$<$ $p_{\rm T}$ $<$8 GeV/$c$), superimposed with the results of a partonic transport model. The contribution of primordial (dashed) and regeneration (continuous line) components is shown, as well as their sum. The area between the curves represent an estimate of the theoretical uncertainty.}
\label{fig:3}
\end{figure}

A deeper insight of the mechanisms at play in Pb-Pb at the LHC can be obtained by studying the centrality dependence of $R_{\rm AA}$ in bins of transverse momentum. If indeed (re)generation mechanisms play a role, their impact should be much more important, for simple kinematic arguments, at low transverse momentum. A larger $R_{\rm AA}$ should then be observed, in particular for central collisions, at low $p_{\rm T}$ with respect to high $p_{\rm T}$.
In Fig.~\ref{fig:3} we separately show the centrality dependence of the 
J/$\psi$ $R_{\rm AA}$ for $0<p_{\rm T}<2$ and $5<p_{\rm T}<8$ GeV/$c$. A larger $R_{\rm AA}$ can indeed be observed at lower $p_{\rm T}$. The results are compared with the calculation of a partonic transport model~\cite{Zha11} which is in good agreement with the data, in particular for central collisions. The contribution of (re)generated J/$\psi$ is shown as continuous lines in the 
plots and indeed is sizeable only at low $p_{\rm T}$.
Similar conclusions can be reached by a direct study of the $p_{\rm T}$ dependence of $R_{\rm AA}$, both centrality integrated and in bins of $N_{\rm part}$~\cite{Sui12,Arn12}.

\begin{figure}[htbp]
\begin{center}
\includegraphics[width=0.48\textwidth]{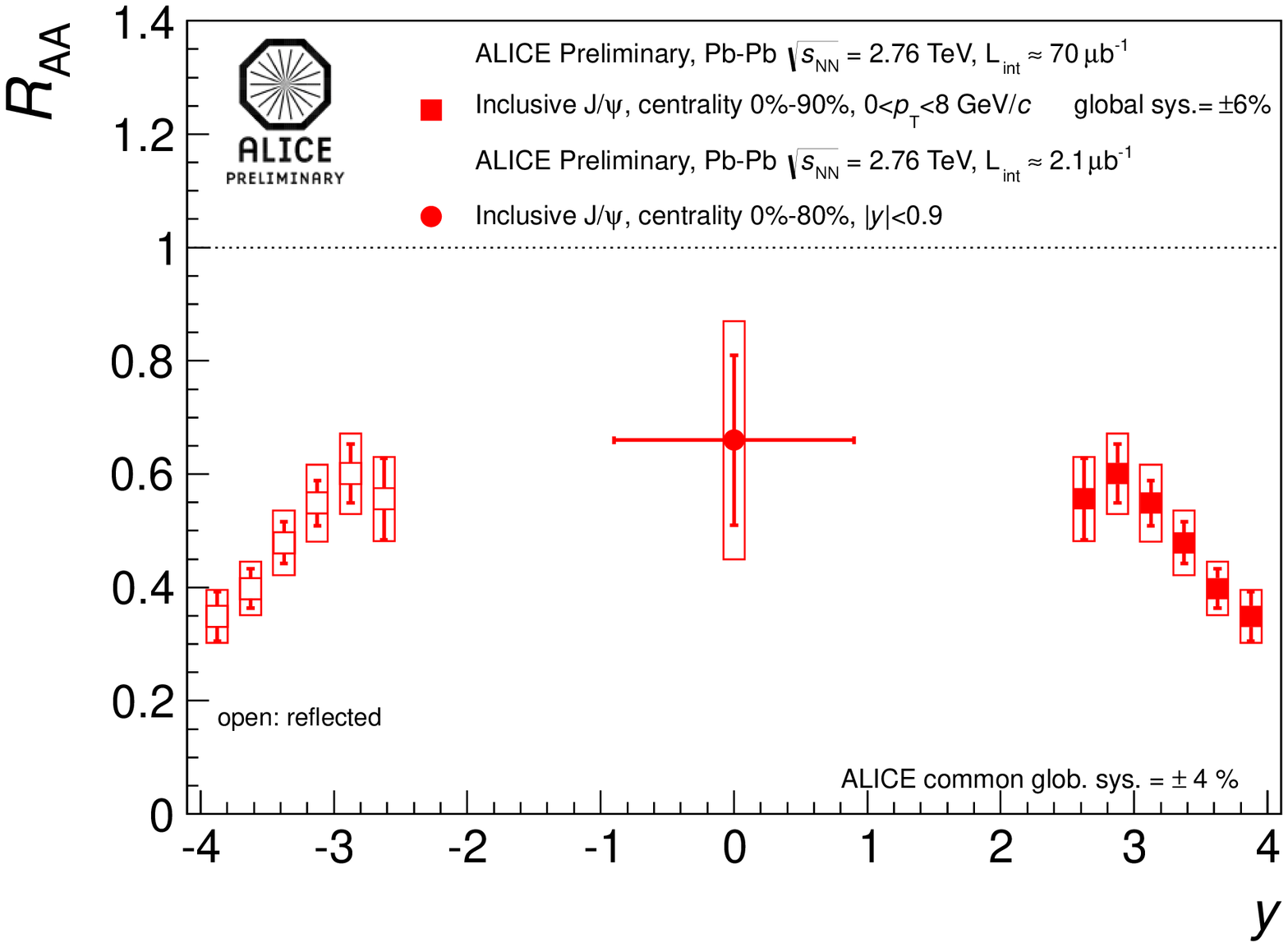}
\includegraphics[width=0.48\textwidth]{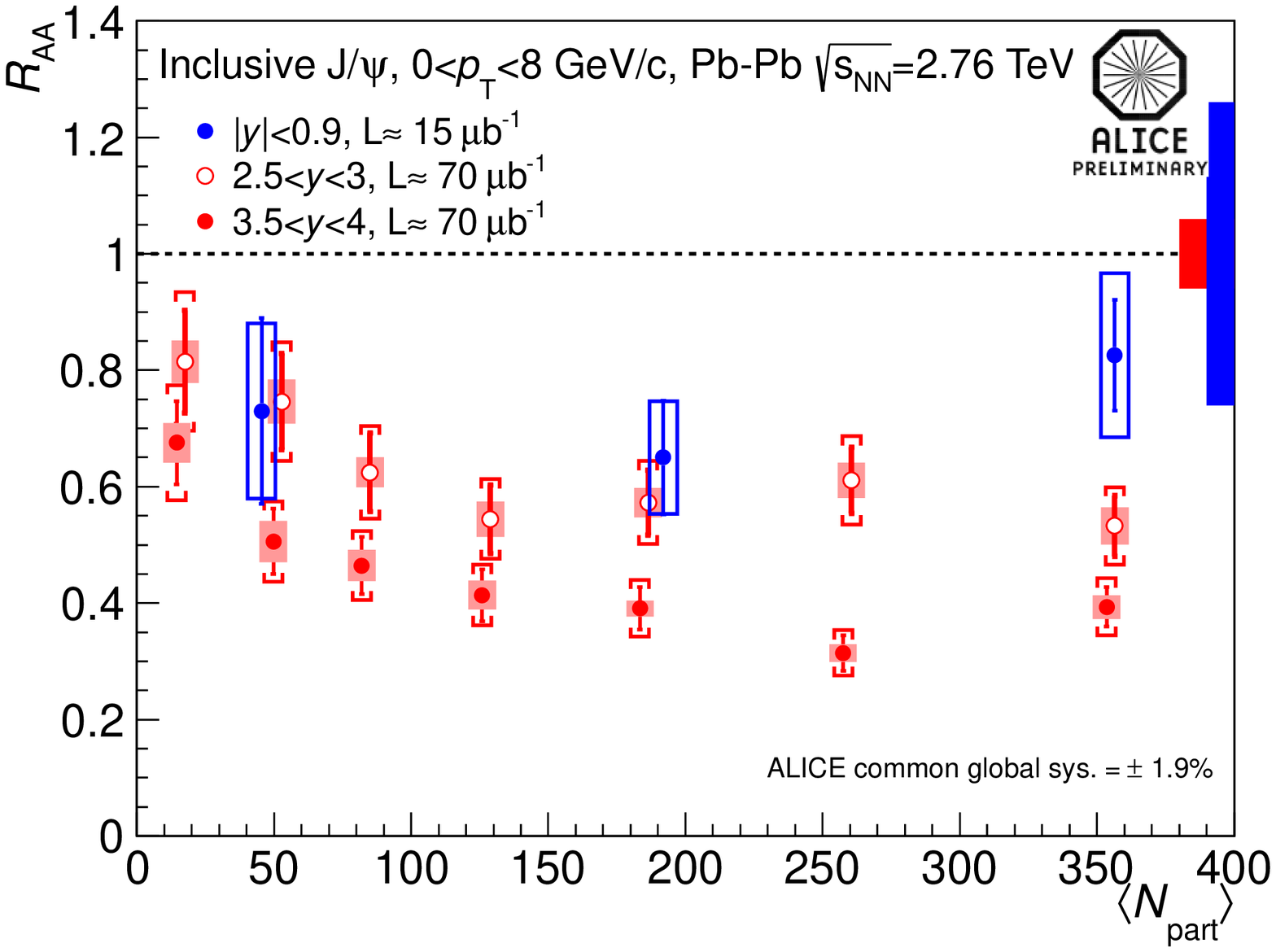}
\end{center}
\vskip -0.5truecm
\caption{Left: $R_{\rm AA}$ vs $y$ for Pb-Pb collisions. 
Global systematic uncertainties are shown in the legends. Right: $R_{\rm AA}$ vs centrality in $y$ bins. For the central rapidity points, the boxes show the uncorrelated systematic uncertainties. For the forward rapidity points the shaded boxes represent uncorrelated systematic uncertainties, while the brackets correspond to uncertainties correlated for that rapidity bin, but uncorrelated between various rapidity bins. Finally, the filled boxes on the right represent global systematic uncertainties, uncorrelated between central and forward rapidity.}
\label{fig:4}
\end{figure}

Other interesting information can be obtained by studying the rapidity dependence of $R_{\rm AA}$. RHIC results showed a significantly larger suppression at forward rapidity~\cite{Ada11}, an observation which lead to many theoretical speculations, mainly connected with the observation of a (re)combination component at central rapidity, where the charmed pair density is larger, or with a different size of cold nuclear matter effects~\cite{Bra11}. In Fig.~\ref{fig:4} (left) we show the centrality integrated $R_{\rm AA}$ vs rapidity. We note that the central rapidity result comes from the lower statistics 2010 Pb-Pb run, since in the high luminosity 2011 data taking the bulk of the statistics was mainly for central events, due to the trigger selection. One can see, within the forward rapidity range, an increase of the suppression moving to larger $y$. In Fig.~\ref{fig:4} (right) we present the centrality dependence of $R_{\rm AA}$ for various rapidity ranges. The tendence to a larger suppression at forward $y$ can be seen at all centralities. Again, the large normalization error (filled boxes) prevents a firm conclusion from the comparison between central (electron) and forward (muon) results. It can be noted that shadowing calculations performed using recent parameterizations (EKS09, nDSg) suggest that the suppression observed at midrapidity could be compatible with a pure nuclear shadowing scenario, while other effects must be invoked in order to reproduce the $R_{\rm AA}$ observed at forward rapidity~\cite{Sui12,Ars12}.
 
Finally, it must be noted that for the moment no data on cold nuclear matter effects on J/$\psi$ exist, at LHC energy. A precise determination of shadowing effects and of the J/$\psi$ break-up probability via interactions with cold nuclear matter (the latter effect might become negligible at the LHC, due to crossing times much smaller than charmonium formation time) can be obtained by studying p-Pb collisions. Such a run will take place at the beginning of 2013 and will help to make the current conclusions from $R_{\rm AA}$ studies more robust.

The differential $R_{\rm AA}$ results previously discussed can be considered as strong hints for an important contribution of (re)generated J/$\psi$ at 
low $p_{\rm T}$ in Pb-Pb collisions at the LHC. An independent confirmation of this hypothesis may come from the study of the elliptic flow of J/$\psi$. If (re)generation effects are sizeable, then the corresponding J/$\psi$ would inherit the flow related to the collective expansion, which is experienced by the charm quarks contained in the fireball. Theoretical models predict in this case a non-zero $v_2$ for the J/$\psi$ at intermediate $p_{\rm T}$~\cite{Liu10}.

\begin{figure}[htbp]
\begin{center}
\includegraphics[width=0.48\textwidth]{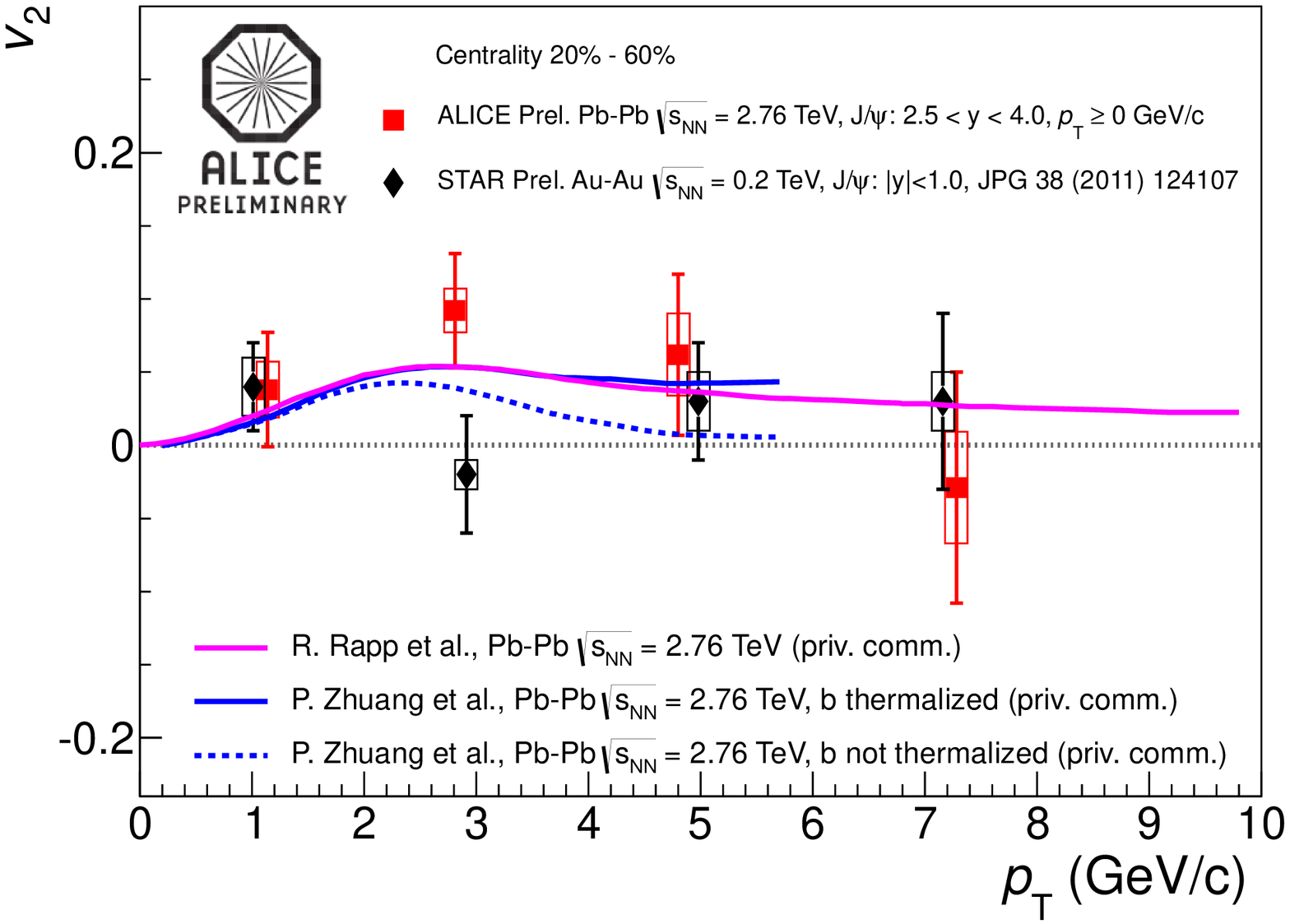}
\includegraphics[width=0.48\textwidth]{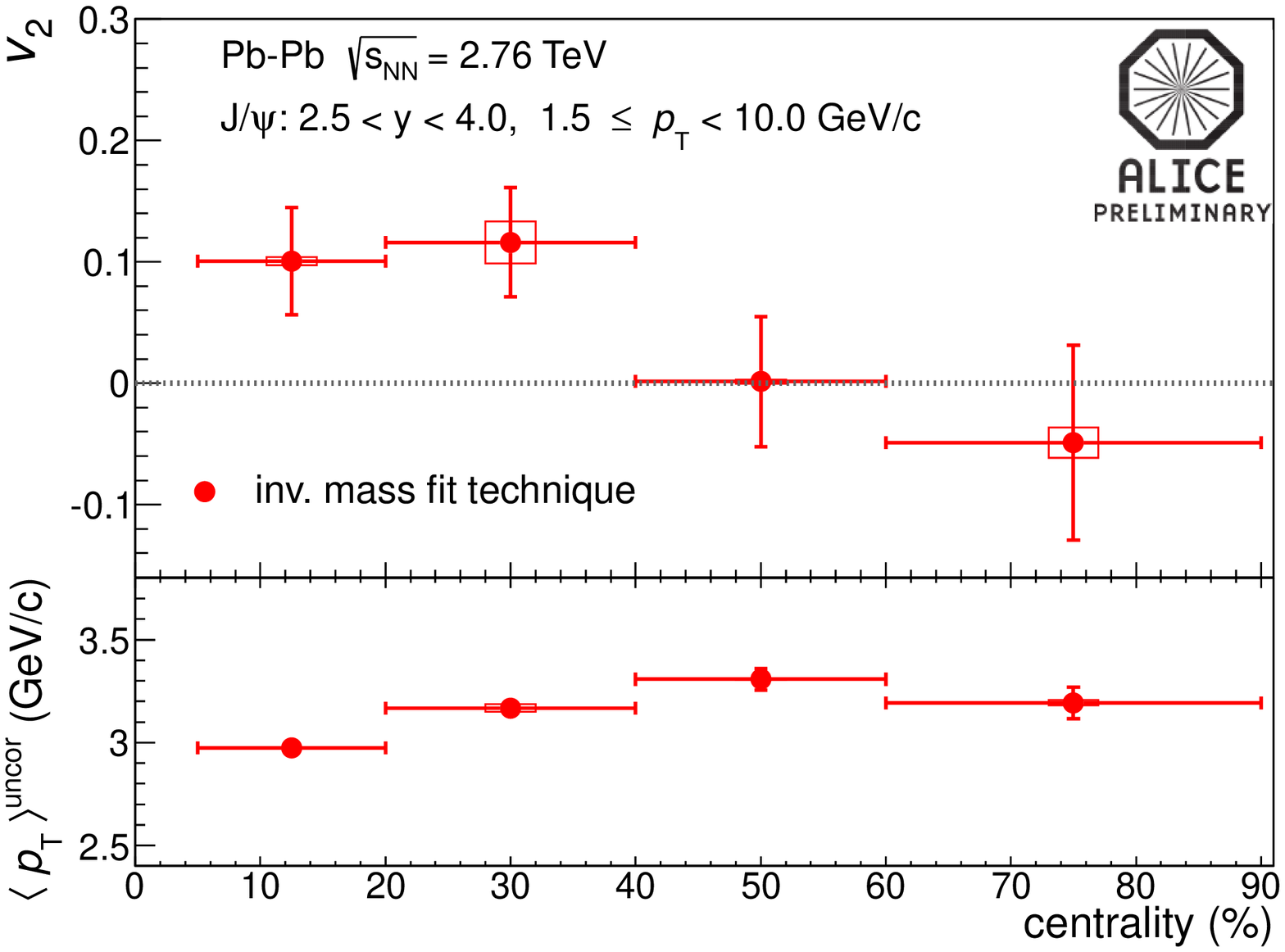}
\end{center}
\vskip -0.5truecm
\caption{Left: the $p_{\rm T}$ dependence of the inclusive J/$\psi$ $v_{\rm 2}$ in $2.5<y<4$, for Pb-Pb collisions (20-60\% centrality). The results are compared to STAR values obtained at central rapidity, in the same centrality range. Calculations from partonic transport models are also shown. Right: the centrality dependence of the inclusive J/$\psi$ $v_{\rm 2}$ in $2.5<y<4$, $1.5<p_{\rm T}<10$ GeV/$c$. The bottom panel shows the average transverse momentum corresponding to each centrality bin. For both plots (left and right panels) the boxes around the points represent the systematic uncertainties.}
\label{fig:5}
\end{figure}

For this study, which was performed on the forward rapidity event sample, methods based on the event plane determination have been used. The signal in the VZERO detector was used for such a purpose, and the usual fit of the J/$\psi$ azimuthal distribution adopted for these studies, ${\rm d}N^{\rm J/\psi}/{\rm d}\Delta\phi = A \times (1+2v^{\rm obs}_2\cos 2\Delta\phi)$ was performed, where $\Delta\phi$ is the azimuthal angle of the J/$\psi$ measured with respect to the event plane.
The $v^{\rm obs}_2$ quantity is then corrected for the event plane resolution, and plotted in Fig.~\ref{fig:5} (left) in four $p_{\rm T}$ bins, using the 20-60\% centrality selection (see also~\cite{Yan12}). Such a binning in $p_{\rm T}$ and centrality was adopted in order to have a more significant comparison with previous preliminary results from the STAR Collaboration~\cite{Tan11}. While STAR results are compatible with zero everywhere, a hint for a non-zero $v_2$ can be appreciated at intermediate $p_{\rm T}$ in the ALICE results. As pointed out above, partonic transport models indeed predict such a kind of behaviour.
In Fig.~\ref{fig:5}(right), the forward rapidity inclusive J/$\psi$ $v_2$ is shown as a function of centrality, in the range $1.5<p_{\rm T}<10$ GeV/$c$. Again, clear hints for non-zero $v_2$ can be appreciated for semi-central collisions. Depending on the adopted binning for the study, the significance for non-zero $v_2$ reaches about 3$\sigma$.

\begin{figure}[htbp]
\begin{center}
\includegraphics[width=0.48\textwidth]{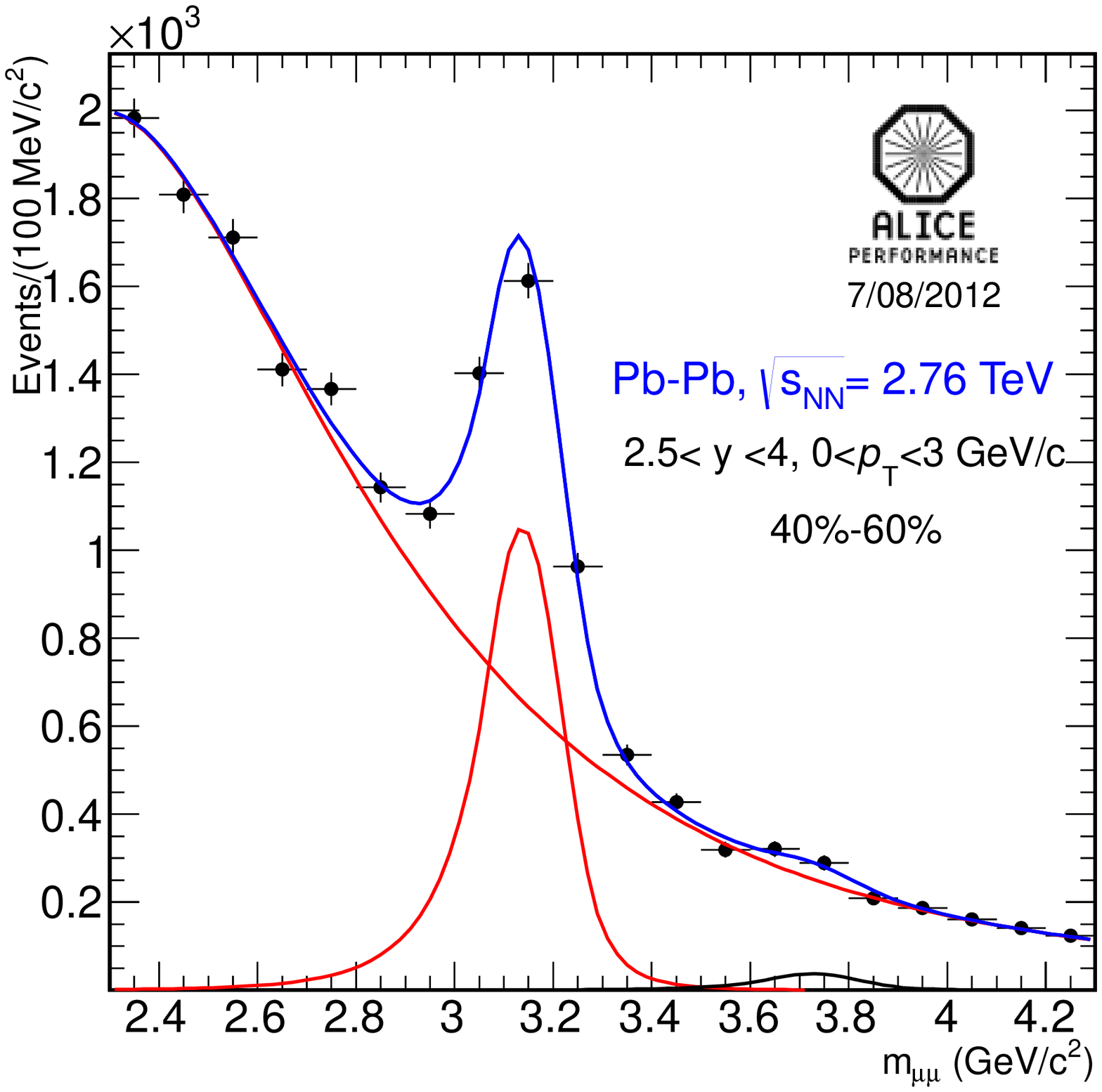}
\includegraphics[width=0.44\textwidth]{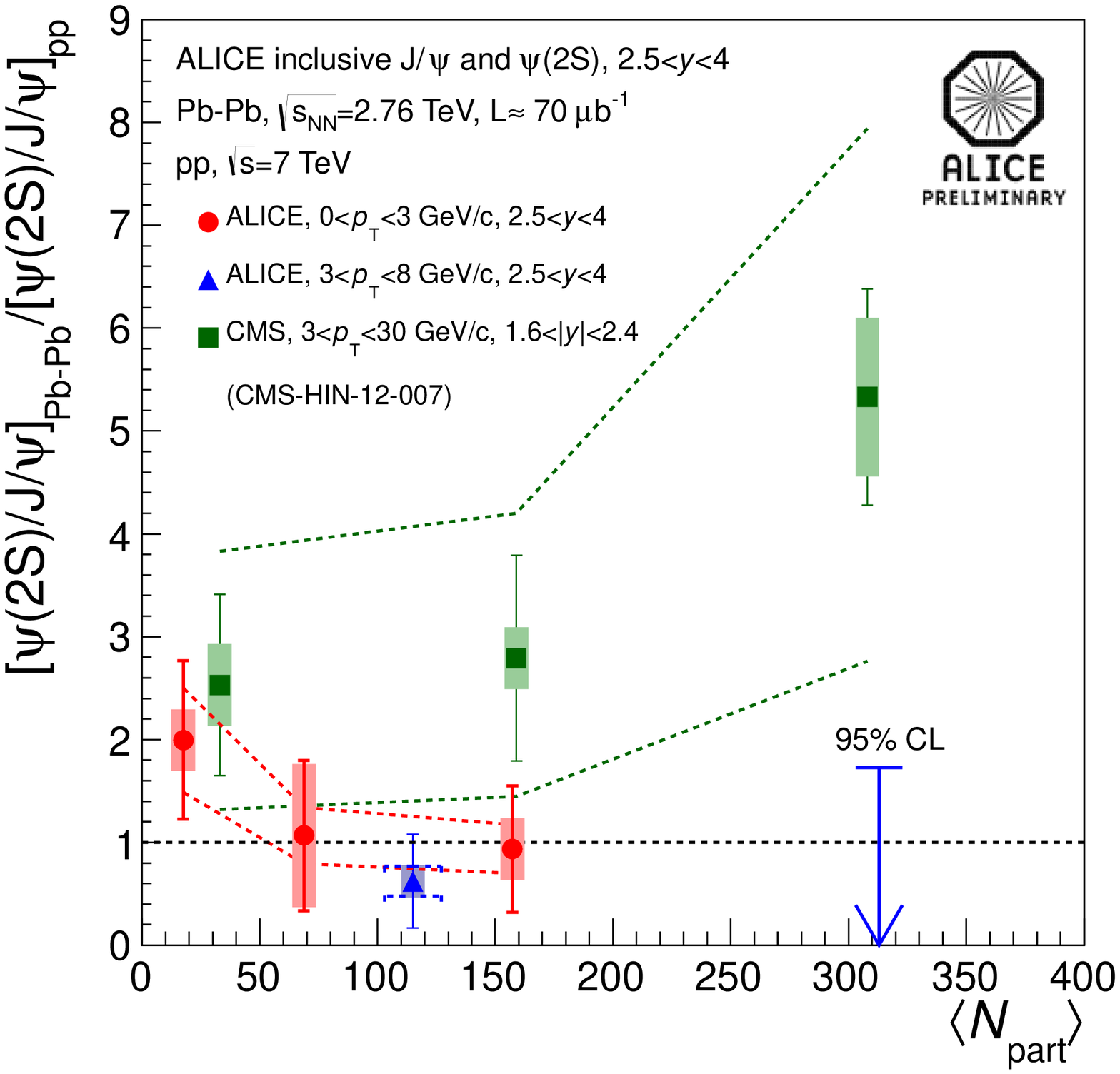}
\end{center}
\vskip -0.5truecm
\caption{Left: the fit to the dimuon invariant mass spectrum for Pb-Pb collisions, for $0<p_{\rm T}<3$ GeV/$c$, in the centrality range 40-60\%.  Right: the centrality dependence of the double ratio $[\psi(2S)/{\rm J}/\psi]_{Pb-Pb}/[\psi(2S)/{\rm J}/\psi]_{pp}$ of the $\psi(2S)$ and J/$\psi$ yields at forward rapidity, compared to results from CMS ($1.6<|y|<2.4$). The dotted lines refer to the pp reference uncertainties, while the shaded areas around the points represent the other systematic uncertainties.}
\label{fig:6}
\end{figure}

Finally, the comparative study of the yields of various charmonium resonances in heavy-ion collisions is known to be an interesting tool to test theoretical models which implement different scenarios (sequential suppression, statistical production). Up to very recently the only results came from SPS energy, where a suppression of the $\psi(2S)$ with respect to the J/$\psi$ was observed by NA50~\cite{Ale07}, increasing with centrality and with hints of a saturation for central events. Very recently, CMS studied the double ratio $\psi(2S)/{\rm J}/\psi$ between Pb-Pb and pp collisions~\cite{Dah12}, observing  values higher than 1 in the region $1.6<|y|<2.4$, $3<p_{\rm T}<30$ GeV/$c$, although with rather large normalization errors, mainly due to the low integrated pp luminosity at $\sqrt{s}=2.76$ TeV. ALICE can explore a contiguous range in rapidity ($2.5<y<4$) and also extend the 
$p_{\rm T}$ reach of this measurement down to zero. In Fig.~\ref{fig:6} (left) the invariant mass spectrum for the region $0<p_{\rm T}<3$ GeV/$c$, corresponding to a 40-60\% selection in centrality, is presented. The result of a fit which includes a $\psi(2S)$ component is also shown, and a clear hint for a signal can be seen, in spite of the rather unfavourable signal over background ratio. In Fig.~\ref{fig:6} (right) we show the values of the double ratio $\psi(2S)/{\rm J}/\psi$ measured by ALICE in two $p_{\rm T}$ ranges. The pp values entering this ratio were calculated from the $\sqrt{s}=7$ TeV data sample, because of the too low integrated luminosity at $\sqrt{s}=2.76$ TeV. The possible $\sqrt{s}$-dependence of the ratio was evaluated from a comparison of CDF, CMS and LHCb results and included ($\sim 15$\%) in the systematic uncertainty. Several other sources of systematic uncertainty cancel in the ratio, the main ones surviving being those on signal extraction and on the choice of the MC inputs for acceptance calculation. For central events, only an upper limit can be quoted for 
$3<p_{\rm T}<8$ GeV/$c$, while background levels are too large to extract a meaningful result in the $0<p_{\rm T}<3$ GeV/$c$ bin. Due to the size of the errors, no firm conclusion on an enhancement or suppression of $\psi(2S)/{\rm J}/\psi$ in Pb-Pb with respect to pp can be drawn. However, our data tend to exclude a large enhancement in central collisions.

\section{Conclusions}

The ALICE Collaboration has performed a study of inclusive J/$\psi$ production in pp and Pb-Pb collisions at the LHC, down to zero $p_{\rm T}$. The main results obtained include a suppression which saturates for semi-central to central events to values larger than the ones observed at RHIC.
Differential studies of $R_{\rm AA}$ vs centrality for various $p_{\rm T}$ bins seem to favour a scenario where (re)combination processes play a sizeable role. The observed hints for a non-zero $v_2$ signal at intermediate $p_{\rm T}$ are in agreement with such a picture. Forthcoming p-A data will help to quantify cold nuclear matter effects and hopefully to sharpen the physics interpretation of the Pb-Pb results.  
 


\begin{thebibliography}{00} 
\bibitem{Mat86} T.~Matsui and H.~Satz, Phys. Lett. {\bf B178} (1986) 416.
\bibitem{ZCo11} Z.~Conesa del Valle et al., Nucl. Phys. {\bf B214} (2011) 3. 
\bibitem{Cap08} A.~Capella, L.~Bravina, E.G.~Ferreiro, A.B.~Kaidalov and E.~Zabrodin, Eur. Phys. J. {\bf C58} (2008) 437.
\bibitem{Vog00} R.~Vogt, Phys. Rev. {\bf C81} (2010) 044903; R.~Vogt, Phys. Rev. {\bf C61} (2000) 035203.
\bibitem{Ale05} B.~Alessandro et al. (NA50 Collaboration), Eur. Phys. J. {\bf C39} (2005) 335.
\bibitem{Arn07} R.~Arnaldi et al. (NA60 Collaboration), Phys. Rev. Lett. {\bf 96} (2007) 132302.
\bibitem{Ada11} A.~Adare et al. (PHENIX Collaboration), Phys. Rev. {\bf C84} (2011) 054912.
\bibitem{Ada12} L.~Adamczyk et al. (STAR Collaboration), arXiv:1208.2736.
\bibitem{Bra11} N.~Brambilla et al., Eur. Phys. J. {\bf C71} (2011) 1534.
\bibitem{Abe12} B. Abelev et al. (ALICE Collaboration), Phys. Rev. Lett. {\bf 109} (2012) 072301.
\bibitem{Zha11} X.~Zhao and R.~Rapp, Nucl. Phys. {\bf A859} (2011) 114.
\bibitem{Liu09} Y.~Liu, Z.~Qu, N.~Xu and P.~Zhuang, Phys. Lett. {\bf B678} (2009) 72.
\bibitem{And11} A.~Andronic, P.~Braun-Munzinger, K.~Redlich and J.~Stachel, J. Phys. {\bf G38} (2011) 124081.
\bibitem{Aam08} K.~Aamodt et al. (ALICE Collaboration), JINST {\bf 3} (2008) S08002.
\bibitem{Ars12} I.~Arsene et al. (ALICE Collaboration), these proceedings.
\bibitem{Aam11} K.~Aamodt et al. (ALICE Collaboration), Phys. Lett. {\bf B704} (2011) 442.
\bibitem{Abe122} B.~Abelev et al. (ALICE Collaboration), arXiv:1203.3641.
\bibitem{Abe12p} B.~Abelev et al. (ALICE Collaboration), Phys. Rev. Lett. {\bf 108} (2012) 082001.
\bibitem{But12} M.~Butenschoen and B.A.~Kniehl, Phys. Rev. Lett. {\bf 108} (2012) 172002.
\bibitem{Abe12m} B.~Abelev et al. (ALICE Collaboration), Phys. Lett. {\bf B712} (2012) 165.
\bibitem{Abe12b} B.~Abelev et al. (ALICE Collaboration), arXiv:1205.5580.
\bibitem{Sui12} C.~Suire et al. (ALICE Collaboration), arXiv:1208.5601.
\bibitem{Arn12} R.~Arnaldi et al. (ALICE Collaboration), these proceedings.
\bibitem{Liu10} Y.~Liu, N.~Xu and P.~Zhuang, Nucl. Phys. {\bf A834} (2010) 317c.
\bibitem{Yan12} H.~Yang et al. (ALICE Collaboration), these proceedings.
\bibitem{Tan11} Z.~Tang et al. (STAR Collaboration), J. Phys. {\bf G38} (2011) 124107.
\bibitem{Ale07} B.~Alessandro et al. (NA50 Collaboration), Eur. Phys. J. {\bf C49} (2007) 559.
\bibitem{Dah12} T.~Dahms et al. (CMS Collaboration), arXiv:1209.3661.
\end{thebibliography}
\end{document}